\documentclass[aps,prl,twocolumn,superscriptaddress]{revtex4}
\usepackage{graphicx}

\def\beq{\begin{equation}}
\def\eeq{\end{equation}}
\def\bea{\begin{eqnarray}}
\def\eea{\end{eqnarray}}

\def\bwt{\begin{widetext}}
\def\ewt{\end{widetext}}

\nocite{*}

\usepackage[hypertex]{hyperref}

\begin{document}

\title{{\it Gravitational Aether} and the thermodynamic solution to the cosmological constant problem}

\author{Niayesh Afshordi}\email{nafshordi@perimeterinstitute.ca}\affiliation{Perimeter Institute
for Theoretical Physics, 31 Caroline St. N., Waterloo, ON, N2L 2Y5,Canada}

\date{\today}
\preprint{astro-ph/yymmnnn}
\begin{abstract}

Cosmological constant problem (in its various versions) is arguably the deepest gap in our understanding of theoretical physics, the solution to which may very likely require revisiting the Einstein theory of gravity. In this letter, I argue that the simplest consistent way to decouple gravity from the vacuum energy (and hence solve the problem) is through the introduction of an incompressible {\it gravitational aether} fluid. The theory then predicts that gravitational constant for radiation is $33\%$ larger than that of non-relativistic matter, which is preferred by most cosmological observations (with the exception of light element abundances), but is not probed by current precision tests of gravity. I also show that slow-roll inflation can happen in this theory, with only minor modifications.  Finally, interpreting gravitational aether as a thermodynamic description of gravity, I propose a finite-temperature correction to the equation of state of gravity, which would explain the present-day acceleration of the cosmic expansion as a consequence of the formation of stellar mass black holes.

\end{abstract}
\maketitle

The cosmological constant problem, along with the nature of quantum gravity, are arguably the deepest and most longstanding gaps in our understanding of theoretical physics. On the one hand, consistent quantization of Einstein theory of gravity is impossible due to non-renormalizable divergences associated with high energy gravitons that run in loops. Attempts to regulate these divergences require introducing new degrees of freedom (such as space-time discreteness) or replacing particles by extended objects (such as strings and non-perturbative higher dimensional branes). Such attempts, even though very fruitful over the past century, have not yet been able to convincingly reproduce the standard cosmological or particle physics models.

On the other hand, in order to explain cosmological observations (or even the mere fact that there {\it are} cosmological observations) the energy density of vacuum fluctuations of well-established quantum field theories should exquisitely cancel that of yet-to-be discovered fields, as well as possible non-perturbative effects to better than $1$ part in $10^{60}$ (see e.g. \cite{Weinberg:1988cp, Polchinski:2006gy} for reviews). The most popular solution to this problem is the proposal that the Universe realizes very different vacuum densities on causally disconnected patches, but the observers can only live in regions hospitable to structure formation and life (known as the anthropic selection), which would choose small vacuum densities \cite{Weinberg:1987dv}. While logically plausible, almost all other predictions of the anthropic solution (i.e. the multi-verse) lie beyond the observable horizon, and thus cannot be tested or falsified. This does not vouch well for a physical theory. Moreover, other physical and cosmological parameters (in particular, the amplitude of density perturbations) could be different in different patches, rendering the anthropic criterion for vacuum density ill-defined \cite{Aguirre:2001zx}.

Alternatively, one may envisage that a modification of Einstein gravity would decouple the vacuum from gravity. This is the route that we will pursue here. A realization of this possibility, known as {\it degravitation} introduces a high-pass filter in the linear theory of gravity, so that gravity is shut off for long wavelength density perturbations (thus degravitating the vacuum) \cite{Dvali:2007kt}. Unfortunately, the only proposed non-linear completion of this theory, known as {\it cascading gravity} \cite{deRham:2007xp,deRham:2007rw} does not as yet have any known cosmological solutions, due to its complexity. Moreover, degravitation cannot explain the coincidence problem, i.e. why the onset of cosmic acceleration was coincident with structure formation in the Universe.

 While both problems (cosmological constant and quantum gravity) involve the interaction of quantum mechanics with the theory of gravity, the proposed solutions to each problem, while often incomplete on their own, also appear completely divorced \footnote{The only possible exception, known to this author, might be populating the string landscape with an anthropic prescription, which could solve both problems within the broad context of string theory. However, like other anthropic solutions, this would suffer from lack of predictibility/falsifibility.}. Here, however I speculate that as the two problems have common ingredients, they might also have a common solution.

 In particular, an action principle, which is {\it only necessary for a consistent quantization}, may also be only existent in terms of the fundamental gravitational degrees of freedom. The effective metric degrees of freedom, observable to us at low energies, may instead have only a {\it thermodynamic} description which lack an effective action and/or unitarity after {\it coarse-graining} the fundamental degrees of freedom. This also would be consistent with the thermodynamic description of general relativity \cite{Jacobson:1995ab}, as well as black hole entropy and its information paradox, since a classical action theory need not have a thermodynamic description.

 A well-known example of a similar situation is the Navier-Stokes equation in fluid mechanics, which gives an effective coarse-grained description of the phase space density of particles. While an action principle can be written for individual particles, the Navier-Stokes equation, coupled with the heat transfer equation, lack an action and only provide a thermodynamic description.

Therefore, from here on, we will abandon the action principle, and instead try to build up a classical thermodynamic theory of gravity consistent with experimental/observational constraints, most important of which is decoupling the quantum vacuum from gravity. Following the language of Jacobson \cite{Jacobson:1995ab}, we will seek a new {\it equation of state} for gravity, as a modification of the Einstein equation.

  As deviations from Einstein gravity in vacuum are already thoroughly tested through solar system/astrophysical observations or terrestrial experiments (see \cite{Will:2001mx} for a review), we will only modify the source of gravity, i.e. the right hand side of the Einstein equation. The most general local covariant modification of the right hand side of the Einstein equation which:

a) is linear in the energy-momentum tensor $T_{\mu\nu}$, as expected at low energies, and

b) is insensitive to the vacuum energy density, $\rho_{\rm vac}$, where
$
T_{\mu\nu} = \rho_{\rm vac}g_{\mu\nu} + {\rm excitations},$

is
\beq
(8\pi G')^{-1} G_{\mu\nu}[g_{\mu\nu}] = T_{\mu\nu} - \frac{1}{4}Tg_{\mu\nu} + ...~ .\label{new}
\eeq
This modified Einstein equation, if self-consistent and in agreement with other experimental bounds on gravity, could potentially constitute a solution to the cosmological constant problem. Notice that Eq. (\ref{new}) is in contrast with unimodular gravity, where both sides of the Einstein equation are modified (and traceless).

The reason we have added ``$...$'' to the right hand side of Eq. (\ref{new}) is that, because of the Bianchi identity, the left hand side of Eq. (\ref{new}) has zero divergence, while the divergence of $T_{\mu\nu}$ vanishes if we assume matter couples to the space-time geometry through the metric $g_{\mu\nu}$. Therefore, as the new term $- \frac{1}{4}Tg_{\mu\nu}$ has a generally non-vanishing divergence, one needs an additional term to make Eq. (\ref{new}) consistent. For this term, we will assume a simple perfect fluid hypothesis, with a fixed equation of state, $\omega'$, i.e.:
\bea
(8\pi G')^{-1} G_{\mu\nu}[g_{\mu\nu}] = T_{\mu\nu} - \frac{1}{4}Tg_{\mu\nu} + T'_{\mu\nu} ,\label{modified}\\
T'_{\mu\nu} = p'\left[(1+\omega'^{-1})u_{\mu}u_{\nu} - g_{\mu\nu}\right],\label{eq:aether}
\eea
which we call {\it gravitational aether} in our framework (borrowing the terminology of \cite{Jacobson:2004ts}). For the right hand side of Eq. (\ref{modified}) to be divergenceless, we then require:
\beq
{{{T'}_{\mu}}^{\nu}}_{;\nu} =\frac{1}{4} T_{,\nu} .\label{aether_cont}
\eeq

While it might seem that we have replaced an unknown (albeit big) number, $\rho_{\rm vac}$, with unknown scalar and vector fields $p'$ and $u_{\mu}$, we now argue that they are dynamically fixed in terms of $T_{\mu\nu}$ via Eq. (\ref{aether_cont}) (although the dependence will be generally non-local). Let us first consider a homogeneous cosmology with a matter component with fixed equation of state $w$. As the matter satisfies the usual continuity equation, we still have $\rho = \rho_0 a^{-3(1+w)}$. Now, Eq. (\ref{aether_cont}) for a similarly homogeneous gravitational aether yields:
\beq
\frac{dp'}{dt} + 3(1+\omega')H p' = -\frac{3\omega'(1+w)(1-3w)H\rho}{4}.\label{aether_hom}
\eeq
 This equation can be easily solved, as a superposition of homogeneous and inhomogeneous solutions. For $\omega' >  w$, the homogeneous solution will decay faster than the inhomogeneous one, and thus the asymptotic solution will be independent of the initial state of the aether field. For the inhomogeneous solution, we can define an effective gravitational constant $G_{\rm eff}$ so that:
 \beq
  H^2+\frac{k}{a^2} = \frac{8\pi G_{\rm eff}}{3}\rho.
 \eeq
  Using Eq. (\ref{aether_hom}) we find
 \beq
 G_{\rm eff}= 1 - \frac{T}{4\rho} + \frac{p'}{\omega'\rho} = \frac{3}{4}(1+w)\left(\omega'-1/3\over \omega' - w\right)G'.\label{geff_1}
 \eeq
  In other words,  the effective $G$ that relates geometry to the matter density $\rho$ in Friedmann equation is now dependent on the equation of state of the dominant energy component of the Universe. For the specific cases of matter era versus radiation era, we find:
  \beq
  \frac{G_N}{G_R} \equiv \frac{G_{\rm eff}(w=0)}{G_{\rm eff}(w=1/3)} = \frac{3}{4}-\frac{1}{4\omega'} < 0.75.
  \eeq
  \begin{figure}
\includegraphics[width=\columnwidth]{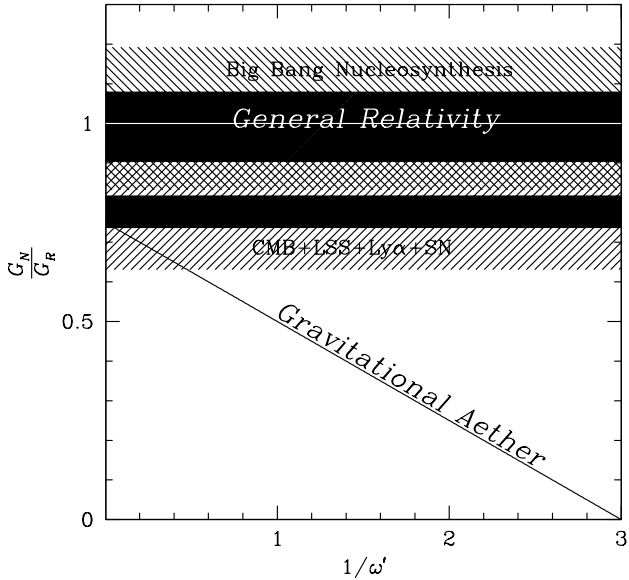}
\caption{The ratio of the Newton's constant for non-relativistic matter (dust), $G_N$, to the effective gravitational constant for radiation $G_R$, as predicted in the {\it gravitational aether} theory with equation of state $\omega'$. The dark (dashed) areas show 68\% (95\%) observational constraints from Big Bang Nucleosynthesis \cite{Cyburt:2004yc} and CMB+LSS+Ly$\alpha$+SNe \cite{Seljak:2006bg}.    } \label{gn_gr}
\end{figure}
  The expansion history in the radiation era depends on the product $G\rho_{\rm rad}$, and is constrained through different observational probes. The constraints are often described as the bound on the effective number of neutrinos $N_{\nu, {\rm eff}}$, which quantifies the total radiation density $\rho_{\rm rad}$. However, assuming only three neutrino specifies, we can translate the constraints to those on $G_{\rm eff}$.  Matching Big Bang Nucleosynthesis predictions with the observed primordial abundances of light elements already requires $G_N/G_R = 0.97 \pm 0.09$ \cite{Cyburt:2004yc}, which is discrepant with any value of $\omega'>0$, at $>2.4\sigma$ level, although it clearly prefers larger values of $\omega'$. However, we should note that observational constraints at lower redshifts, and particularly combination of Ly-$\alpha$ forest and CMB observations \cite{Seljak:2006bg} yield $G_{N}/G_{R} = 0.73 \pm 0.04$ which prefers our prediction to that of General Relativity (i.e. $G_N = G_R$), as long as $\omega' \gtrsim 5$ (see Fig. \ref{gn_gr}). Future CMB observations by Planck satellite, as well as ground-based observatories are expected to improve this constraint dramatically over the next five years, and thus confirm or rule out this prediction.

  Motivated by these constraints, we will assume $\omega' \gg 1$ for the rest of the analysis, corresponding to a nearly incompressible fluid, which is sourced by $\nabla T$ (Eq. \ref{aether_cont}). This is also desirable, as in the $\omega'\rightarrow \infty$ limit (or the {\it cuscuton} fluid) there is no superluminal propagation of information \cite{Afshordi:2006ad}. Moreover, in the incompressible limit, there will be no energy loss to aether sound waves, which is tightly constrained from the binary pulsar observations (e.g. see \cite{Will:2001mx}). In the non-relativistic regime, the corresponding  continuity and Euler equations become:
  \bea
  \dot{p'}+ (1+\omega')p' {\bf\nabla\cdot u} = \frac{\omega'}{4} \dot{T}, \label{eq:cont}\\
  (1+\omega'^{-1})p'(\dot{\bf u} + \nabla \phi) = -{\bf \nabla}\left(p'+\frac{1}{4}T\right),
  \eea
where $\dot{} \equiv \frac{\partial}{\partial t} + {\bf u\cdot\nabla}$, is the usual Lagrangian (or comoving) time derivative, and $\phi$ is the Newtonian potential. We next notice that for the aether fluid to remain non-relativistic, the effective pressure in the Euler equation should remain small:
\beq
p'+\frac{1}{4} T = p'{\cal O}(u^2,\phi) + {\rm const.}
\eeq
For example, for a static spherical star in an asymptotically flat space-time, assuming that $p' \rightarrow 0$ at infinity, we can integrate the Euler equation to find:
\beq p'(r)+ \frac{1}{4} T(r) = \frac{1}{4}(1+\omega'^{-1}) \int^r_{\infty} T(r') \frac{d\phi(r')}{dr'}dr'+{\cal O}(T\phi^2).\eeq
As the speed of sound for the aether fluid perturbations is simply $c_s = \omega'^{1/2} > 1$ it is not surprising that the perturbations around a static background remain stable:
\beq
\frac{\partial^2 T{\bf u}}{\partial t^2} = \omega'  {\bf \nabla \nabla}\cdot(T{\bf u}). \label{eq:wave}\eeq
Therefore, after the sound waves leave the system (which happens infinitely fast in the incompressible limit $\omega' \rightarrow \infty$), the aether simply follows the non-relativistic matter. The departure from this behavior can be estimated by contrasting the continuity equation of aether (\ref{eq:cont}) with that of matter:
\beq
|{\bf u}_{\rm aether}-{\bf u}_{\rm matter}|  \sim w{\bf u}_{\rm matter},\label{eq:wu}
\eeq
which is small, as long as the gravitating objects have non-relativistic {\it internal} pressure.

In contrast,  the transverse (or vorticity) modes do not propagate in Eq. (\ref{eq:wave}). In principle, similar to the ordinary vector modes in cosmology, these modes are suppressed during the cosmological expansion, but can be sourced during the non-linear collapse phase. For these modes, we have:
$
{\bf u}_{\rm vor.} = \frac{\nabla \times {\bf A}}{T},
$
where ${\bf A}$ is fixed by the boundary conditions of the non-relativistic region. As a result, within dense objects (such as planets or stars), which are the dominant source of gravity in all precision tests of general relativity, the vortical modes nearly vanish in the rest frame of the objects, as $T\simeq \rho$  is several orders of magnitude larger than the surrounding environment. We thus conclude that the incompressible ($\omega' \gg 1$) aether is almost perfectly dragged along with the dense objects with non-relativistic internal pressure. In this regime, the right hand side of the modified Einstein equation (\ref{modified}-\ref{eq:aether}) will take a perfect fluid form: $T_{\rm eff}^{\mu\nu} =(\rho_{\rm eff}+p_{\rm eff})u^{\mu}u^\nu -p_{\rm eff} g^{\mu\nu}$, with:
\beq
\rho_{\rm eff} = \frac{3}{4} (\rho+p),~
p_{\rm eff} = \frac{\dot{p}}{\dot{\rho}}\rho_{\rm eff}, \label{rho_eff}
\eeq
where we used Bianchi identity to substitute for $p'$. In other words, for a slowly varying equation of state $w = p/\rho \simeq \dot{p}/\dot{\rho}$, the effect of gravitational aether is simply to renormalize the effective gravitational constant:
\beq
G_{\rm eff} = \left[1+w+{\cal O}(w^2{\bf u}^2)\right] G_{N}, \label{Geff}
\eeq
where $G_{N} = 3G'/4$, and the size of corrections is determined by the velocity offset between aether and matter (\ref{eq:wu}) \footnote{A Friedmann equation with $G_{\rm eff}$ similar to Eq. (\ref{Geff}) was pursued in \cite{Carroll:2001zy}, although they could not find a concrete model with this exact behavior.}.
It is interesting to notice that, despite all the exquisite precision tests of general relativity (often described in terms of constraints on Parametrized Post-Newtonian (PPN) parameters), a possible anomalous dependence on pressure (the PPN parameter $\zeta_4$) has never been probed independently \cite{Will:2001mx}. In particular, the theoretical expectation that $\zeta_4$ must be related to other PPN parameters (Eq. 39 in \cite{Will:2001mx}) is not realized in the gravitational aether theory, where $\zeta_4 =1/3$ (using Eq. \ref{Geff}), while other PPN parameters vanish.

Let us now consider other cosmological implications of the gravitational aether theory. As the equation of state is negligible in the matter era, the cosmological dynamics is indistinguishable from general relativity. However, the transition from radiation to matter era is delayed as radiation now gravitates $33\%$ stronger than non-relativistic matter. As we argued above (see Fig. \ref{gn_gr}), this is actually preferred by most cosmological observations \cite{Seljak:2006bg}. In the radiation era, $T=0$ and thus again, the dynamics is indistinguishable from general relativity.

One may wonder if we can realize a successful inflationary scenario in the gravitational aether theory, especially since inflation is (often) driven by the (near) vacuum energy of a scalar field. However, we should note that, since inflation should come to an end, the energy density cannot be exactly constant. Combining Eqs. (\ref{rho_eff}) and field equation for a scalar field $\varphi$ with the potential $V(\varphi)$, we see that there still exists a phase of slow-roll inflation if:
\beq
-\frac{2\dot{H}}{3H^2} = 1+w_{\rm eff} = 1+\frac{\dot{p}}{\dot{\rho}} \simeq  \frac{M_p V''(\varphi)}{\sqrt{3} |V'(\varphi)|} \ll 1,
\eeq
where $M_p$ is the reduced Planck mass ($M^{-2}_p \equiv 8\pi G_N = 6\pi G'$). Notice that the expansion rate now depends on $\dot{\varphi}$, rather than $V(\varphi)$ in the modified Friedmann equation:
\beq
3M^2_p H^2= \dot{\varphi}^2 \simeq \frac{M_p}{\sqrt{3}}|V'(\varphi)|,\label{friedmann}
\eeq
The power spectrum of scalar quantum fluctuations can be found similar to ordinary inflation:
\bea
{\cal P}_{\zeta} = \frac{H^4}{4\pi^2\dot{\varphi}^2} = \frac{H^2}{12\pi^2M^2_p} \simeq \frac{|V'(\varphi)|}{36\sqrt{3}\pi^2M_p^3},\\
n_s-1 \equiv \frac{d\ln P_{\zeta}(k)}{d\ln k} \simeq -\sqrt{3}M_p \frac{V''(\varphi)}{|V'(\varphi)|},
\eea
with the notable absence of $\epsilon = M^2_p (V'/V)^2/2$ in the denominator of ${\cal P}_{\zeta}$ (in contrast with ordinary inflation), as we have used the modified Friedmann equation (\ref{friedmann}).
However, without an action for the gravity sector, we cannot make a prediction for the amplitude of gravity waves generated via quantum vacuum fluctuations. I postpone a more comprehensive study of the inflationary/ekpyrotic scenarios to future publications.

A final piece of the puzzle is the discovery of recent cosmic acceleration \cite{Riess:1998cb,Perlmutter:1998np} which is now confirmed by several complementary lines of evidence, and has rejuvenated the interest in the cosmological constant problem over the last decade. Now that we have decoupled gravity from $\rho_{\rm vac}$, how can we explain the present-day cosmic acceleration, which is often interpreted as evidence for a small, but non-vanishing $\rho_{\rm vac}$ in the context of general relativity? Similar to inflation, we can imagine a ``quintessence''-like scalar potential that could drive a present-day inflationary phase. However, similar to ordinary quintessence, it should have a very small mass $\sim 10^{-33} {\rm eV}$, which is not natural in the presence of any coupling to standard model fields. Moreover, it cannot explain the coincidence of structure formation and cosmic acceleration.

Let us speculate a more attractive possibility: Following our starting point, i.e. a thermodynamic description of gravity, we should expect finite-temperature corrections in the equation of state of gravity (or gravitational aether) (Eqs. \ref{modified}-\ref{eq:aether}). In particular, the coincidence of the formation of black holes, with well-defined gravitational horizon temperatures, $T_{H} = M^2_p/M_{BH}$, with the onset of cosmic acceleration may be providing us with an empirical measure of these corrections. In particular, a correction of the form:
\beq
T'_{\mu\nu}= p'(u_{\mu}u_{\nu}-g_{\mu\nu}) + \alpha M_p T^3_{H} g_{\mu\nu},\label{finite_temp}
\eeq
with $\alpha = {\cal O}(1)$ would exactly reproduce the observed vacuum energy density, $\Omega_{\Lambda} \simeq 0.7$, if:
\beq
M_{BH} \simeq (20 ~M_{\odot})~ \alpha^{1/3},
\eeq
 which is extremely suggestive, as this is exactly the mass range expected for black holes produced in supernova explosions at the end of the lifetime of massive stars. Nevertheless, we should note that this might also just be a numerical coincidence. Moreover, in the absence of an understanding of the fundamental gravitational degrees of freedom, we cannot justify why finite-temperature corrections to the aether energy should grow as $T^3_{H}$ (rather than e.g. $T_H^2$ or $T^4_H$), or even whether the aether could reach thermal equilibrium.

 More optimistically, we could view Eq. (\ref{finite_temp}) as an empirical determination of the equation of state of gravity, with the crucial assumption of {\it no fine-tuning} of the dimensionless parameters of the theory. Similar to the development of the kinetic theory of gases starting from the observed ideal gas law, one may decide to view Eq. (\ref{finite_temp}) as a guideline for the development of a more fundamental theory of gravity, which would then serve as a significantly more testable alternative to the anthropic solutions of the cosmological constant problem.


I am indebted to Ghazal Geshnizjani, Justin Khoury, Kazunory Kohri, Lee Smolin, Rafael Sorkin, and in particular to Andrew Tolley for invaluable discussions and comments.
This work is supported by Perimeter Institute for Theoretical Physics.
 Research at Perimeter Institute is supported by the Government of Canada through Industry Canada and by
 the Province of Ontario through the Ministry of Research \& Innovation.

\bibliographystyle{utcaps_na2}
\bibliography{aether}

\end{document}